\documentclass[runningheads]{llncs}
\usepackage{graphicx}
\usepackage[misc,geometry]{ifsym}
\usepackage{amsfonts}
\usepackage{nicefrac} 

\usepackage{floatrow}
\newfloatcommand{capbtabbox}{table}[][\FBwidth] 

\usepackage{amsmath,amsfonts,bm}

\def\eqref#1{equation~\ref{#1}}

\def\1{\bm{1}}

\def\vx{{\bm{x}}}

\def\mA{{\bm{A}}}

\def\mE{{\bm{E}}}

\def\mH{{\bm{H}}}
\def\mI{{\bm{I}}}

\def\mX{{\bm{X}}}

\DeclareMathAlphabet{\mathsfit}{\encodingdefault}{\sfdefault}{m}{sl}
\SetMathAlphabet{\mathsfit}{bold}{\encodingdefault}{\sfdefault}{bx}{n}

\newcommand{\ptitle}[1]{\vspace{1mm}\noindent{\bf #1.}}
\begin{document}
%
\title{Adversarial Robustness of Probabilistic Network Embedding for Link Prediction}
\titlerunning{Adversarial Robustness of Link Prediction}

\author{Xi Chen\inst{1} (\Letter) \and
Bo Kang\inst{1}\and
Jefrey Lijffijt\inst{1}\and
Tijl De Bie\inst{1}}
\authorrunning{X. Chen et al.}
%
\institute{IDLab, Department of Electronics and Information Systems, Ghent University, Technologiepark-Zwijnaarde 122, 9052 Ghent, Belgium \\
\email{\{firstname.lastname\}@ugent.be}}
\maketitle              
\begin{abstract}
In today's networked society, many real-world problems can be formalized as predicting links in networks, such as Facebook friendship suggestions, e-commerce recommendations, and the prediction of scientific collaborations in citation networks. 
Increasingly often, link prediction problem is tackled by means of network embedding methods, owing to their state-of-the-art performance. 
However, these methods lack transparency when compared to simpler baselines, and as a result their robustness against adversarial attacks is a possible point of concern: could one or a few small adversarial modifications to the network have a large impact on the link prediction performance when using a network embedding model? 
Prior research has already investigated adversarial robustness for network embedding models, focused on classification at the node and graph level. 
Robustness with respect to the link prediction downstream task, on the other hand, has been explored much less.

This paper contributes to filling this gap, by studying adversarial robustness of Conditional Network Embedding (CNE), a state-of-the-art probabilistic network embedding model, for link prediction. 
More specifically, given CNE and a network, we measure the sensitivity of the link predictions of the model to small adversarial perturbations of the network, namely changes of the link status of a node pair. Thus, our approach allows one to identify the links and non-links in the network that are most vulnerable to such perturbations, for further investigation by an analyst. 
We analyze the characteristics of the most and least sensitive perturbations, and empirically confirm that our approach not only succeeds in identifying the most vulnerable links and non-links, but also that it does so in a time-efficient manner thanks to an effective approximation.
\end{abstract}

\section{Introduction \label{sec:intro}}
Networks are used to model entities and the relations among them, so they are capable of describing a wide range of data in real world, 
such as social networks, citation networks, and networks of neurons.
The recently proposed Network Embedding (NE) methods can be used to learn representations of the non-iid network data such that networks are transformed into the tabular form.
The tabular data can then be fed to solve several network tasks, such as visualization, node classification, recommendation, and link prediction.
We focus on link prediction that aims to predict future or currently missing links~\cite{liben2007link} as it has been widely applied in our lives.
Examples include Facebook friendship suggestions, Netflix recommendations, predictions of protein-protein interactions, etc.

Many traditional link prediction approaches have been proposed~\cite{martinez2016survey}, but the task is tackled increasingly often by the NE methods due to their state-of-the-art performance~\cite{mara2020benchmarking}.
However, the NE methods lack transparency, e.g., Graph Neural Networks (GNNs)~\cite{gori2005new}, when compared to simpler baselines.
Thus, similar to many other machine learning algorithms~\cite{goodfellow2014explaining}, they could be vulnerable to adversarial attacks.
It has been shown that simple imperceptible changes of the node attribute or the network topology can result in wrongly predicted node labels, especially for GNNs~\cite{zugner2018adversarial,dai2018adversarial}.
Meanwhile, adversarial attacks are easy to be found in our daily {\it online} lives, such as in recommender systems~\cite{zhang2020practical,liu2021adversarial,yang2017fake}.

Robustness of NE methods for link prediction is important.
Attacking link prediction methods can be used to hide sensitive links, while defending can help identify the interactions hidden intentionally, e.g., important connections in crime networks.
Moreover, as links in online social networks represent the information sources and exposures, from the dynamic perspective, manipulations of network topology can be used to affect the formation of public opinions on certain topics, e.g., via exposing a targeted group of individuals to certain information sources, which is risky.
The problem we want to investigate is: {\it Could one or a few small adversarial modifications to the network topology have a large impact on the link prediction performance when using a network embedding model?}

Existing adversarial robustness studies for NE methods mainly consider classification at the node and graph level, 
which investigates whether the labels will be wrongly predicted due to adversarial perturbations.
It includes semi-supervised node classification~\cite{zugner2018adversarial,zugner2020adversarial,zugner2019adversarial,xu2019topology,zhu2019robust,wu2019adversarial,feng2019graph,bojchevski2019certifiable,zugner2020certifiable,zugner2019certifiable,tang2020transferring}, 
and graph classification~\cite{dai2018adversarial,ma2019attacking,jin2020certified}.
Only a few works consider the link-level task~\cite{lin2020adversarial,chen2020link,bojchevski2019adversarial,dai2019adversarial}, leaving robustness of NE methods for link prediction insufficiently explored.

To fill the gap, we study the adversarial robustness of Conditional Network Embedding (CNE)~\cite{kang2018conditional} for the link prediction task.
CNE is a state-of-the-art probabilistic NE model that preserves the first-order proximity, of which the objective function is expressed analytically.
Therefore, it provides mathematically principled explainability~\cite{kang2019explaine}.
Moreover, comparing to other NE models, such as those based on random walks~\cite{perozzi2014deepwalk,grover2016node2vec}, 
CNE is more friendly to link prediction because the link probabilities follow directly from the model so there is no need to further train a classifier for links with the node embeddings.
However, there has been no study on the adversarial robustness of CNE for link prediction. 

In our work, we consider only the network topology as input, meaning that there is no node attribute.
More specifically, given CNE and a network, we measure the sensitivity of the link predictions of the model to small adversarial perturbations of the network, i.e., the changes of the link status of a node pair.
The sensitivity is measured as the impact of the perturbation on the link predictions.
Intuitively, we quantify the impact as the KL-divergence between the two link probability distributions learned by the model from the clean and the corrupted network through re-training.
While the re-training can be expensive, we develop effective and efficient approximations based on the gradient information, which is similar to the computation of the regularizer in Virtual Adversarial Training (VAT)~\cite{miyato2018virtual}.
Our main contributions are:
\begin{itemize}
\item We propose to study the adversarial robustness of a probabilistic network embedding model CNE for link prediction;
\item Our approach allows us to identify the links and non-links in the network that are most vulnerable to adversarial perturbations for further investigation;
\item With two case studies, we explain the robustness of CNE for link prediction through (a) illustrating how structural perturbations affect the link predictions; (b) analyzing the characteristics of the most and least sensitive perturbations, providing insights for adversarial learning for link prediction.
\item We show empirically that our gradient-based approximation for measuring the sensitivity of CNE for link prediction to small structural perturbations is not only time-efficient but also significantly effective.
\end{itemize}

\section{Related Work \label{sec:related_work}} 
Robustness in machine learning means that a method can function correctly with erroneous inputs~\cite{ieee1990ieee}.
The input data may contain random noise embedded, or adversarial noise injected intentionally.
The topic became a point of concern when the addition of noise to an image, which is imperceptible to human eyes, resulted in a totally irrelevant prediction label~\cite{goodfellow2014explaining}.
Robustness of models against noisy input has been investigated in many works~\cite{mirzasoleiman2020coresets,zheng2020robust,dai2018adversarialNE}, 
while adversarial robustness usually deals with the worst-case perturbations on the input data.

Network tasks at the node, link, and graph level are increasingly done by network embedding methods, which include shallow models and GNNs~\cite{liunetwork}.
Shallow models either preserve the proximities between nodes (e.g., DeepWalk~\cite{perozzi2014deepwalk}, LINE~\cite{tang2015line}, and node2vec~\cite{grover2016node2vec}) or factorize matrices containing graph information~\cite{wang2017community,qiu2018network} to effectively represent the nodes as vectors.
GNNs use deep structure to extract node features by iteratively aggregating their neighborhood information, e.g., Graph Convolutional Networks (GCNs)~\cite{kipf2017semi} and GraphSAGE~\cite{hamilton2017inductive}.

Adversarial learning for networks includes three types of studies: attack, defense, and certifiable robustness~\cite{sun2018adversarial,jin2020adversarial,chen2020survey}.
Adversarial attacks aim to maximally degrade the model performance through perturbing the input data, which includes the modification of node attributes or changes of the network topology.
Examples of attacking strategies for GNNs include the non-gradient based NETTACK~\cite{zugner2018adversarial}, Mettack using meta learning~\cite{zugner2019adversarial}, SL-S2V with reinforcement learning~\cite{dai2018adversarial}, and attacks by rewiring for graph classification~\cite{ma2019attacking}.
The defense strategies are designed to protect the models from being attacked in many different ways, e.g., by detecting and recovering the perturbations~\cite{wu2019adversarial}, applying adversarial training~\cite{goodfellow2014explaining} to resist the worst-case perturbation~\cite{feng2019graph}, or transferring the ability to discriminate adversarial edges from exploring clean graphs~\cite{tang2020transferring}.
Certifiable robustness is similar in essence to adversarial defense, but it focuses on guarantee the reliability of the predictions under certain amounts of attacks.
The first provable robustness for GNNs was proposed to certify if a node label will be changed under a bounded attack on node attributes~\cite{zugner2019certifiable}, and later a similar certificate for structural attack was proposed~\cite{zugner2020certifiable}.
There are also robustness certifications for graph classification~\cite{jin2020certified,gao2020certified} and community detection~\cite{jia2020certified}.
The most popular combination is GNNs for node or graph classification, while the link-level tasks has been explored much less.

Early studies on robustness for link-level tasks usually target traditional link prediction approaches.
That includes link prediction attacks that aim to solve specific problems in the social context, e.g., to hide relationships~\cite{fard2015neighborhood,waniek2019hide} or to disguise communities~\cite{waniek2018hiding},
and works that restrict the perturbation type to only adding or only deleting edges~\cite{zhou2019attacking,zhou2019adversarial,yu2019target}, which could result in less efficient attacks or defenses.
The robustness for NE based link prediction is much less investigated than classification, and is considered more often as a way to evaluate the robustness of the NE method, such as in~\cite{pan2018adversarially,bojchevski2019adversarial,sun2018data}.
To the best of our knowledge, there are only two works on adversarial attacks for link prediction based on NE: one targeting the GNN-based SEAL~\cite{zhang2018link} with structural perturbations and one targeting GCN with iterative gradient attack~\cite{chen2020link}.

\section{Preliminaries \label{sec:preliminaries}}
In this section, we provide the preliminaries of our work, 
including the notations,
the probabilistic network embedding model CNE that we use for link prediction, 
and the virtual adversarial training method to which the our idea is similar.

\subsection{Link Prediction with Probabilistic Network Embedding}
Network embedding methods map nodes in a network onto a lower dimensional space as real vectors or distributions, and we work with the former type.
Given a network $G=(V, E)$, where $V$ and $E$ are the node and edge set, respectively, 
a network embedding model finds a mapping $f: V \to \mathbb{R}^d$ for all nodes as $\mX = [\vx_1, \vx_2, ... \vx_n]^T \in \mathbb{R}^{n \times d}$. 
Those embeddings $\mX$ can be used to visualize the network in the $d$-dimensional space; 
classify nodes based on the similarity between vector pairs;
and predict link probabilities between any node pair.

To do link prediction, a network embedding model requires a function $g$ of vectors $\vx_i$ and $\vx_j$ to calculate the probability of nodes $i$ and $j$ being linked.
This can be done by training a classifier with the links and non-links, or the function follows naturally from the model.
Conditional Network Embedding (CNE) is the probabilistic model on which our work is based, and of which the function $g$ directly follows~\cite{kang2018conditional}.
Suppose there is an undirected network $G=(V, E)$ with its adjacency matrix $\mA$, 
where $a_{ij} = 1$ if $(i,j) \in E$ and $0$ otherwise, 
CNE finds an optimal embedding $\mX^*$ that maximizes the probability of the graph conditioned on that embedding.
It maximizes its objective function:
\begin{equation}
P(G | \mX) = \prod_{(i, j) \in E} P(a_{ij} = 1| \mX) \prod_{(k, l) \notin E} P(a_{kl} = 0 | \mX). \label{eq:cne}
\end{equation}
To guarantee that the connected nodes are embedded closer and otherwise farther, the method uses two half normal distributions for the distance $d_{ij}$ between nodes $i$ and $j$ conditioned on their connectivity.
By optimizing the objective in Eq.~(\ref{eq:cne}), CNE finds the most informative embedding $\mX^*$ and the probability distribution $P(G|\mX^*)$ that defines the link predictor $g(\vx_i, \vx_j) = P(a_{ij} = 1| \mX^*)$.

Many network embedding methods purely map nodes into vectors of lower dimensions and focus on node classification, such the random-walk based ones~\cite{perozzi2014deepwalk,grover2016node2vec,tang2015line} and GCNs~\cite{kipf2017semi}.
Those methods require an extra step to measure the similarities between the pairs of node embeddings for link prediction.
Comparing to them, CNE is a better option for link prediction.
Moreover, CNE provides good explainability for link predictions as $g$ can be expressed analytically~\cite{kang2019explaine}.

\subsection{Virtual Adversarial Attack}
Adversarial training achieved great performance for the supervised classification problem~\cite{goodfellow2014explaining},
and virtual adversarial training (VAT) is better for the semi-supervised setting~\cite{miyato2018virtual}.
By identifying the most sensitive `virtual' direction for the classifier, VAT uses regularization to smooth the output distribution.
The regularization term is based on the virtual adversarial loss of possible local perturbations on the input data point.
Let $x \in \mathbb{R}^d$ and $y \in Q$ denote the input data vector of dimension $d$ and the output label in the space of $Q$, respectively.
The labeled data is defined as $\mathcal{D}_l = \left \{ x^{(n)}_l, y^{(n)}_l | n = 1, ..., N_l\right \}$, the unlabeled data as $\mathcal{D}_{ul} = \left \{ x^{(m)}_{ul} | m = 1, ..., N_{ul} \right \}$, and the output distribution as $p(y|x, \theta)$ parametrized by $\theta$.
To quantify the influence of any local perturbation on $x_*$ (either $x_l$ or $x_{ul}$), VAT has the Local Distribution Smoothness (LDS),
\begin{align}
\mathrm{LDS} (x_*, \theta) &:= D\left [ p(y|x_*, \hat{\theta}), p(y|x_* + r_{vadv}, \theta)\right ] \label{eq:LDS}\\
 r_{vadv} &:= \mathrm{argmax}_{r; ||r||_2 \leq \epsilon}  D\left [ p(y|x_*,\hat{\theta}), p(y|x_* + r, \theta)\right ], \label{eq: r_vadv}
\end{align}
where $D$ can be any non-negative function that measures the divergence between two distributions, 
and $p(y|x, \hat{\theta})$ is the current estimate of the true output distribution $q(y|x)$.
The regularization term is the average LDS for all data points.

Although VAT was designed for classification with tabular data, the idea of it is essentially similar to our work, i.e., we both quantify the influence of local virtual adversarial perturbations.
For us, that is the link status of a node pair.
As we have not yet included the training with a regularization term in this work, we now focus on finding the $r_{vadv}$ in Eq.~(\ref{eq: r_vadv}).
That is to identify the most sensitive perturbations that will change the link probabilities the most.

\section{Quantifying the Sensitivity to Small Perturbations \label{sec:method}}
With the preliminaries, we now formally introduce the specific problem we study in this paper.
That is, to investigate if there is any small perturbations to the network that have large impact on the link prediction performance.
The small perturbations we look into are the edge flips, which represent either the deletion of an existing edge or the addition of a non-edge.
It means that we do not restrict the structural perturbations to merely addition or merely deletion of edges.

Intuitively, that impact of any small virtual adversarial perturbation can be measured by re-training the model.
But re-training, namely re-embedding the network using CNE, can be computationally expensive. 
Therefore, we also investigate on approximating the impact both practically with incremental partial re-embedding, and theoretically with the gradient information.

\subsection{Problem Statement and Re-Embedding (RE)}
The study of the adversarial robustness for link prediction involves identifying the worst-case perturbations on the network topology, 
namely the changes of the network topology that influence the link prediction results the most.
For imperceptibility, we focus on the small structural perturbation of individual edge flip in this work.
Thus, our specific problem is defined as
\begin{problem}[Impact of a structural perturbation]\label{prob:impact_of_attack}
Given a network $G = (V, E)$, 
a network embedding model, 
how can we measure the impact of each edge flip in the input network on the link prediction results of the model?
\end{problem}

Intuitively, the impact can be measured by assuming the edge flip as a virtual attack, flip the edge and retrain the model with the virtually corrupted network, after which we know how serious the attack is.
That means we train CNE with the clean graph $G = (V, E)$ to obtain the link probability distribution $P^* = P(G|\mX^*(\mA))$.
After flipping one edge, we get the corrupted graph $G' = (V, E')$, retrain the model, and obtain a different link probability $Q^* = Q(G'|\mX^*(\mA'))$.
Then we measure the impact of the edge flip as the KL-divergence between $P^*$ and $Q^*$.
In this way, we also know how the small perturbation changes the node embeddings, which helps explain the influence of the virtual attack.

If the virtual edge flip is on node pair $(i, j)$, 
$a'_{ij} = 1 - a_{ij}$ where $a_{ij}$ is the corresponding entry in the adjacency matrix of the clean graph $\mA$ and $a'_{ij}$ of the corrupted graph $\mA'$.
Re-embedding $G'$ with CNE results in probability $Q^*(i, j)$,
then the impact of flipping $(i, j)$, which we consider as the sensitivity of the model to the perturbation on that node pair, denoted as $s(i, j)$, is:
\begin{equation}
s(i, j) = KL\left [ P^*||Q^*(i, j) \right ]. \label{eq:sensitivity_ij}
\end{equation}
Measured practically, this KL-divergence is the actual impact for each possible edge flip on the predictions.
The optimal embeddings $\mX^*(\mA)$ and $\mX^*(\mA')$ not only explain the influenced link predictions but also exhibit the result of the flip.

Ranking the node pairs in the network by the sensitivity measure for all node pairs allows us to identify the most and least sensitive links and non-links for further investigation.
However, re-embedding the entire network can be computationally expensive, especially for large networks.
The sensitivity measure can be approximated both empirically and theoretically, and we will show how this can be done in the rest of this section.

\subsection{Incremental Partial Re-Embedding (IPRE)}
Empirically, one way to decrease the computational cost is to incrementally re-embed only the two corresponding nodes of the flipped edge.
In this case, our assumption is that the embeddings of all nodes except the two connecting the flipped edge (i.e., node $i$ and $j$) will stay unchanged since the perturbation is small and local.
We call it Incremental Partial Re-Embedding (IPRE), which allows only the changes of $\vx_i$ and $\vx_j$ if $(i, j)$ is flipped.
It means that the impact of the small perturbation on the link probabilities is restricted within the one-hop neighborhood of the two nodes, resulting in the changed link predictions between node $i$ and $j$ with the rest of the nodes.
The definition of the impact in Eq~(\ref{eq:sensitivity_ij}) stills holds and only the $i$th and $j$th columns and rows in the link probability matrix have non-zero values.
Comparing to RE, IPRE turns out to be a faster and effective approximation, which we will show with experiments.

\subsection{Theoretical Approximation of the KL-Divergence}
Incrementally re-embedding only the two nodes of the flipped edge is faster but it is still re-training of the model.
Although our input is non-iid, in contrast to the tabular data used in VAT~\cite{miyato2018virtual}, we can form our problem as in Eq.~(\ref{eq:vat_lp}), 
of which the solution is the most sensitive structural perturbation for link prediction.
\begin{equation}
\Delta \mA := \mathrm{argmax}_{\Delta \mA; ||\Delta \mA|| = 2} KL \left [  P(G|\mX^*(\hat{\mA})), P(G|\mX^*(\hat{\mA} + \Delta \mA) )\right ]. \label{eq:vat_lp}
\end{equation}

CNE has its link probability distribution expressed analytically, 
so the impact of changing the link status of node pair $(i, j)$,  represented by the KL-divergence in Eq~(\ref{eq:sensitivity_ij}) can be approximated theoretically.
Given the clean graph $G$, CNE learns the optimal link probability distribution $P^* = P(G | \mX^*(\mA))$ whose entry is $P_{kl}^* = P(a_{kl} = 1 | \mX^*)$.
Let $Q^*(i, j)$ be the optimal link probability distribution of the corrupted graph $G'$ with only $(i, j)$ flipped from the clean graph.
The impact of the flip $s(i, j)$ can be decomposed as,
\begin{equation}
s(i, j) = KL\left [P^*||Q^*(i, j) \right ] = \sum \left [ p \log \frac{p}{q} + (1-p) \log \frac{1-p}{1-q} \right ], \label{eq:decompose_sij}
\end{equation}
where $p$ and $q$ are entries of $P^*$ and $Q^*(i, j)$ respectively. 
We can approximate $s(i, j)$ at $G$, or equivalently, at $P^*$, as $G$ is close to $G'$ thus $P^*$ is close to $Q^*(i, j)$.

The first-order approximation of $s(i, j)$ is a constant because at $G$ its gradient $\frac{\partial KL\left [P^*||Q^*(i, j) \right ]}{\partial a_{ij}} =0$,
so we turn to the second-order approximation in Eq.~(\ref{eq:second_order_approx}), which, evaluated at $G$, is $\tilde{s}(i,j)$ in Eq.~(\ref{eq:kl_approx}).
That requires the gradient of each link probability w.r.t the edge flip, i.e., $\frac{\partial p}{\partial a_{ij}} = \frac{ \partial P^*_{kl}}{\partial a_{ij}}$.
Now we will show how to compute it with CNE.
\begin{align}
s(i, j) \approx &\frac{\partial KL\left [P^*||Q^*(i, j) \right ] }{\partial a_{ij}} \Delta \mA + \frac{1}{2} \frac{\partial^2 KL\left [P^*||Q^*(i, j) \right ]) }{\partial a_{ij}^2} \Delta \mA^2, \label{eq:second_order_approx} \\
\tilde{s}(i,j) = &\frac{1}{2} \sum \frac{1}{p \left ( 1-p\right )} \left [ \frac{\partial p}{\partial a_{ij}} \right ]^2. \label{eq:kl_approx}
\end{align}

\ptitle{The gradient}
At the graph level, the gradient of a link probability $P^*_{kl}$ 
for node pair $(k, l)$ w.r.t the input graph $\mA$ is
$\frac{ \partial P^*_{kl}}{\partial \mA} = \frac{\partial P^*_{kl}}{\partial \mX^*(\mA)} \frac{\partial \mX^*(\mA)}{\partial \mA}$.
While at the node pair level, the gradient of $P^*_{kl}$ w.r.t. $a_{ij}$ is 
\begin{align}
\frac{ \partial P^*_{kl}}{\partial a_{ij}}  &= \frac{\partial P^*_{kl}}{\partial \vx^*(\mA)} \frac{\partial \vx^*(\mA)}{\partial a_{ij}}\\
&= \vx^{*T} (\mA) \mE_{kl} \mE_{kl}^T \left [ \frac{ -\mH }{\gamma^2 P^*_{kl} (1-P^*_{kl})} \right ]^{-1} \mE_{ij} \mE_{ij}^T \vx^*(\mA), \label{eq:exact_grad}
\end{align}
where for clearer presentation we flatten the matrix $\mX$ to a vector $\vx$ that is $nd \times 1$, 
$\mE_{kl}$ is a column block matrix consisting of $n$ $d \times d$ blocks where the $k$-th and $l$-th block are positive and negative identity matrix $\mI$ and $-\mI$ of the right size respectively and $0$s elsewhere,
and $\mH$ is the full Hessian below
\begin{equation*}
\mH = \gamma \sum_{u \neq v}\left [  (P^*_{uv} - a_{uv}) \mE_{uv} \mE_{uv}^T  - \gamma P^*_{uv} (1 - P^*_{uv}) \mE_{uv} \mE_{uv}^T \vx^*(\mA) \vx^{*T}(\mA)\mE_{uv} \mE_{uv}^T\right ]. 
\end{equation*}

The gradient reflects the fact that the change of a link status in the network influences the embeddings $\vx^*$, and then the impact is transferred through $\vx^*$ to the link probabilities of the entire graph.
In other words, if an important relation (in a relatively small network) is perturbed, it could cause large changes in many $P^*_{kl}$s, deviating them from their predicted values with the clean graph.

The gradient in Eq.~(\ref{eq:exact_grad}) is exact and measures the impact all over the network.
However, the computation of the inverse of the full Hessian can be expensive when the network size is large.
But fortunately, $\mH$ can be well approximated with its diagonal blocks~\cite{kang2019explaine}, which are of size $d \times d$ each block.
So we can approximate the impact of individual edge flip with $\tilde{s}(i,j)$ at a very low cost using
\begin{equation}
\frac{ \partial P^*_{kl}}{\partial a_{ki}}  = (\vx^*_k - \vx^*_l)^T  \left [ \frac{ -\mH_k }{ \gamma^2 P^*_{kl} (1-P^*_{kl})} \right ]^{-1} (\vx^*_k - \vx^*_i), 
\end{equation}
where $\mH_k = \gamma \sum_{l: l \neq k}\left [  (P^*_{kl} - a_{kl}) \mI  - \gamma P^*_{kl} (1 - P^*_{kl}) (\vx^*_k - \vx^*_l)(\vx^*_k - \vx^*_l)^T\right ]$ is the $k$th diagonal block of $\mH$.
Here $P^*_{kl}$ is assumed to be influenced only by $\vx_k$ and $\vx_l$, thus only the edge flips involving node $k$ or $l$ will result in non-zero gradient for $P^*_{kl}$.
It essentially corresponds to IPRE, where only the attacked nodes are allowed to move in the embedding space.
In fact, as the network size grows, local perturbations are not likely to spread the influence broadly.
We will show empirically this theoretical approximation is both efficient and effective.

\section{Experiments\label{sec:experiments}}
For the purpose of evaluating our work, we first focus on illustrating the robustness of CNE for link prediction with two case studies, 
using two networks of relatively small sizes.
Then we evaluate the approximated sensitivity for node pairs on larger networks.
The research questions we want to investigate are:
\begin{itemize}
\item How to understand the sensitivity of CNE to an edge flip for link prediction?
\item What are the characteristics of the most and least sensitive perturbations for link prediction using CNE?
\item What are the quality and the runtime performance of the approximations?
\end{itemize}

\ptitle{Data}
The data we use includes six real world networks of varying sizes.
{\bf Karate} is a social network of 34 members in a university karate club, which has 78 friendship connections~\cite{zachary1977information}.
{\bf Polbooks} network describes 441 Amazon co-purchasing relations among 105 books about US politics~\cite{adamic2005political}.
{\bf C.elegans} is a neural network of the nematode C.elegans with 297 neurons linked by 2148 synapses~\cite{watts1998collective}.
{\bf USAir} is a transportation network of 332 airports as nodes and 2126 airlines connecting them as links~\cite{handcock2003statnet}.
{\bf MP} is the largest connected part of a Twitter friendship network for the Members of Parliament (MP) in the UK during April 2019, having 567 nodes and 49631 edges~\cite{chen2021alpine}.
{\bf Polblogs} is a network with 1222 political blogs as nodes and 16714 hyperlinks as undirected edges, which is the largest connected part of the US political blogs network from~\cite{adamic2005political}.

\ptitle{Setup}
We do not have train-test split, because we want to measure the sensitivity of {\it all} link probabilities of CNE to {\it all} small perturbations of the network.
The CNE parameters are $\sigma_2 = 2$, $d=2$ for the case studies, $d = 8$ for evaluating the approximation quality, learning rate is $0.2$, $\mathrm{max\_iter}=2k$, and $\mathrm{ftol}=1e-7$.

\subsection{Case Studies}
The first two research questions will be answered with the case studies on Karate and Polbooks, 
which are relatively small thus can be visualized clearly.
Both networks also have ground-truth communities, which contributes to our analysis.
With Karate, we show how the small perturbations influence link probabilities via node embeddings.
On Polbooks, we analyze the characteristics of the most and least sensitive perturbations.
Note that we use the dimension $2$ for both the visualization of CNE embeddings and the calculation of the sensitively.

\ptitle{Karate}
To show the process of attacking CNE link prediction on Karate, 
we illustrate and analyze how the most sensitive edge deletion and addition affect the model in predicting links.
With the RE approach, we measure the model sensitivity to single edge flip and find the top 5 sensitive perturbations in Table~\ref{tab:case_karate}.
The most sensitive deletion of link $(1, 12)$ disconnects the network, and we do not consider this type of perturbation in our work because it is obvious and easy to be detected.
We see the other top sensitive perturbations are all cross-community, 
and we pick node pairs $(1, 32)$ and $(6, 30)$ for further study.

\begin{table}[hbt!]
\begin{floatrow}
\capbtabbox{
\begin{tabular}{|c|c|c|c|c|}
\hline
Rank & Node Pair & s(i, j) & A{[}i, j{]} & Community? \\ \hline
1    & (1, 12)   & 12.30   & 1           & within     \\ \hline
2    & (1, 32)   & 2.52    & 1           & cross      \\ \hline
3    & (20, 34)  & 1.96    & 1           & cross      \\ \hline
4    & (6, 30)   & 1.75    & 0           & cross      \\ \hline
5    & (7, 30)   & 1.75    & 0           & cross      \\ \hline
\end{tabular}}
{
  \caption{The Top 5 Sensitive Perturbations}
  \label{tab:case_karate}
}
\capbtabbox{
\begin{tabular}{|c|c|c|c|}
\hline
          & RE     & IPRE   & Approx  \\ \hline
Polbooks  & 0.889  & 0.117  & 0.00012 \\ \hline
C.elegans & 2.819  & 0.568  & 0.00045 \\ \hline
USAir     & 6.206  & 0.781  & 0.00043 \\ \hline
MP        & 8.539  & 2.289  & 0.00116 \\ \hline
Polblogs  & 45.456 & 27.648 & 0.00124 \\ \hline
\end{tabular}
}{
  \caption{Runtime in seconds}
   \label{tab:runtime}
}
\end{floatrow}
\end{table}

\begin{figure}[hbt!]
\centering
\includegraphics[width=\columnwidth]{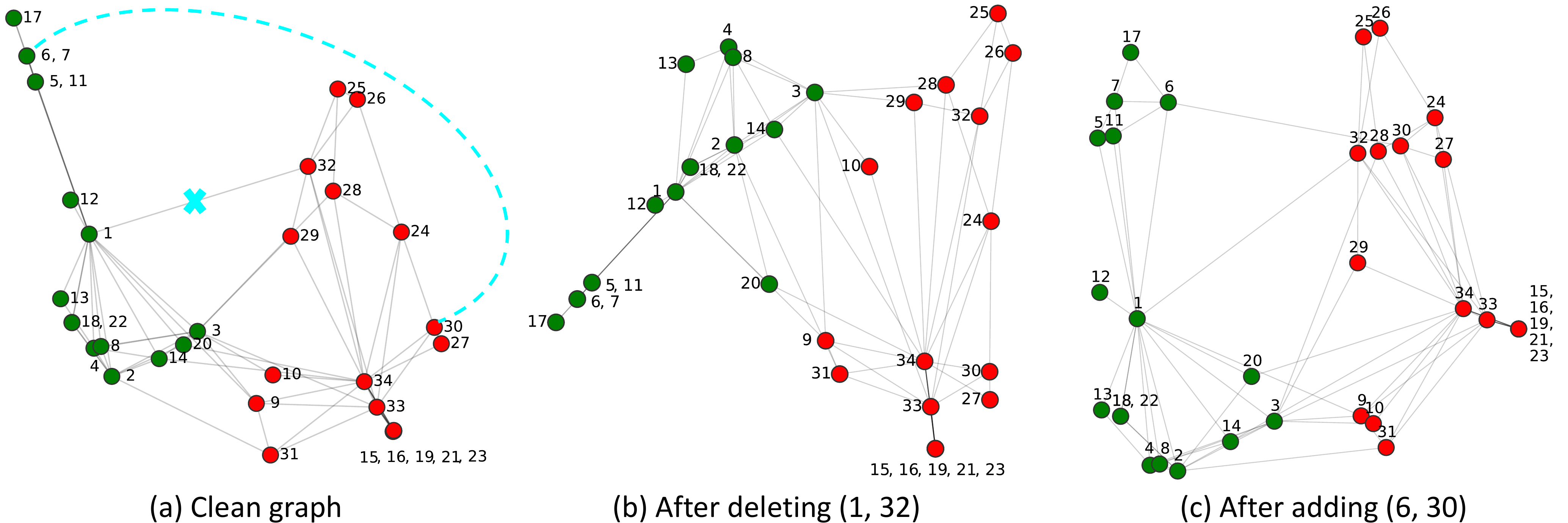}
\vspace{-5mm}
\caption{Case study on Karate with the most sensitive perturbations. \label{fig:case_karate}}
\end{figure}

Fig.~\ref{fig:case_karate} shows the CNE embeddings of the clean Karate and the perturbed graphs,
where the communities are differentiated with green and red color.
CNE embeddings might have nodes overlap when $d=2$, such as node $6$ and $7$, because they have the same neighbors, but this will not be a problem if $d$ is higher.

The deletion of edge $(1, 32)$ is marked with a cross in Fig.~\ref{fig:case_karate} (a), after which the changed node embeddings are shown in Fig.~\ref{fig:case_karate} (b).
Although being rotated, the relative locations of the nodes change a lot, especially node $1$, $32$, and those in the boundary between the communities, e.g., node $3$ and $10$.
Node $1$ is pushed away from the red nodes, and as the center of the green nodes, it plays an essential role in affecting many other link probabilities.
Comparing to other cross-community edges, $(1, 32)$ is the most sensitive because both nodes have each other as the only cross-community link.
So the deletion largely decreases the probability of their neighbors connecting to the other community.
Moreover, node $1$ has a high degree.
Therefore, it makes sense that this is the most sensitive edge deletion.

The addition of edge $(6, 30)$ is marked as a dashed arc in Fig.~\ref{fig:case_karate} (a), and the case is similar for $(7, 30)$.
Adding the edge changes the node locations as shown in Fig.~\ref{fig:case_karate} (c).
The distant tail in green that ends with node $17$ moves closer to the red community.
Note that both node $6$ and $30$ had only the within-community links before the perturbation.
Even though their degrees are not very high, the added edge changes the probabilities of many cross-community links from almost zero to some degree of existence, pulling nodes to the other community.

\ptitle{Polbooks}
Polbooks has three types of political books, which are liberal (L), neutral (N), and conservative (C), marked with colors red, purple, and blue, respectively.
Shown in Table~\ref{tab:case_polbooks} are the most and least sensitive perturbations,
where the left column are the Top 2 deletions and the middle and right columns are the top 5 additions.
We do so as real networks are usually sparse.
The rank is based on the sensitivity measure, thus the non-sensitive perturbations are ranked bottom (i.e., 5460).
Then we will mark the those perturbations in the CNE embeddings, for edge deletions and additions separately.

\begin{table}[hbt!]
\centering
\caption{The Top Sensitive and Non-Sensitive Perturbations}
\vspace{-1mm}
\label{tab:case_polbooks}
\resizebox{\columnwidth}{!}{
\begin{tabular}{|c|c|c|c|c|c|c|c|c|c|c|c|c|c|}
\hline
\multicolumn{4}{|c|}{Edge Deletion - S}     &  & \multicolumn{4}{c|}{Edge Addition - S} &  & \multicolumn{4}{c|}{Edge Addition - Non-S} \\ \hline
Rank   & Node Pair  & s(i, j)  & Community  &  & Rank & Node Pair & s(i, j) & Community &  & Rank  & Node Pair  & s(i, j)  & Community  \\ \hline
1    & (46, 102) & 16.91 & N-L &  & 2 & (3, 98)  & 15.53 & C-L &  & 5458 & (37, 39) & 0.035 & C-C \\ \hline
15   & (7, 58)   & 14.64 & N-C &  & 3 & (3, 87)  & 15.42 & C-L &  & 5454 & (8, 47)  & 0.036 & C-C \\ \hline
\multicolumn{4}{|c|}{Edge Deletion - Non-S} &  & 4    & (28, 33)  & 14.98   & N-C       &  & 5451  & (33, 35)   & 0.038    & C-C        \\ \hline
5460 & (72, 75)  & 0.033 & L-L &  & 5 & (25, 98) & 14.96 & C-L &  & 5449 & (30, 71) & 0.039 & L-L \\ \hline
5459 & (8, 12)   & 0.034 & C-C &  & 6 & (25, 91) & 14.92 & C-L &  & 5438 & (66, 75) & 0.042 & L-L \\ \hline
\end{tabular}
}
\end{table}

\begin{figure}[hbt!]
\centering
\includegraphics[width=0.57\columnwidth]{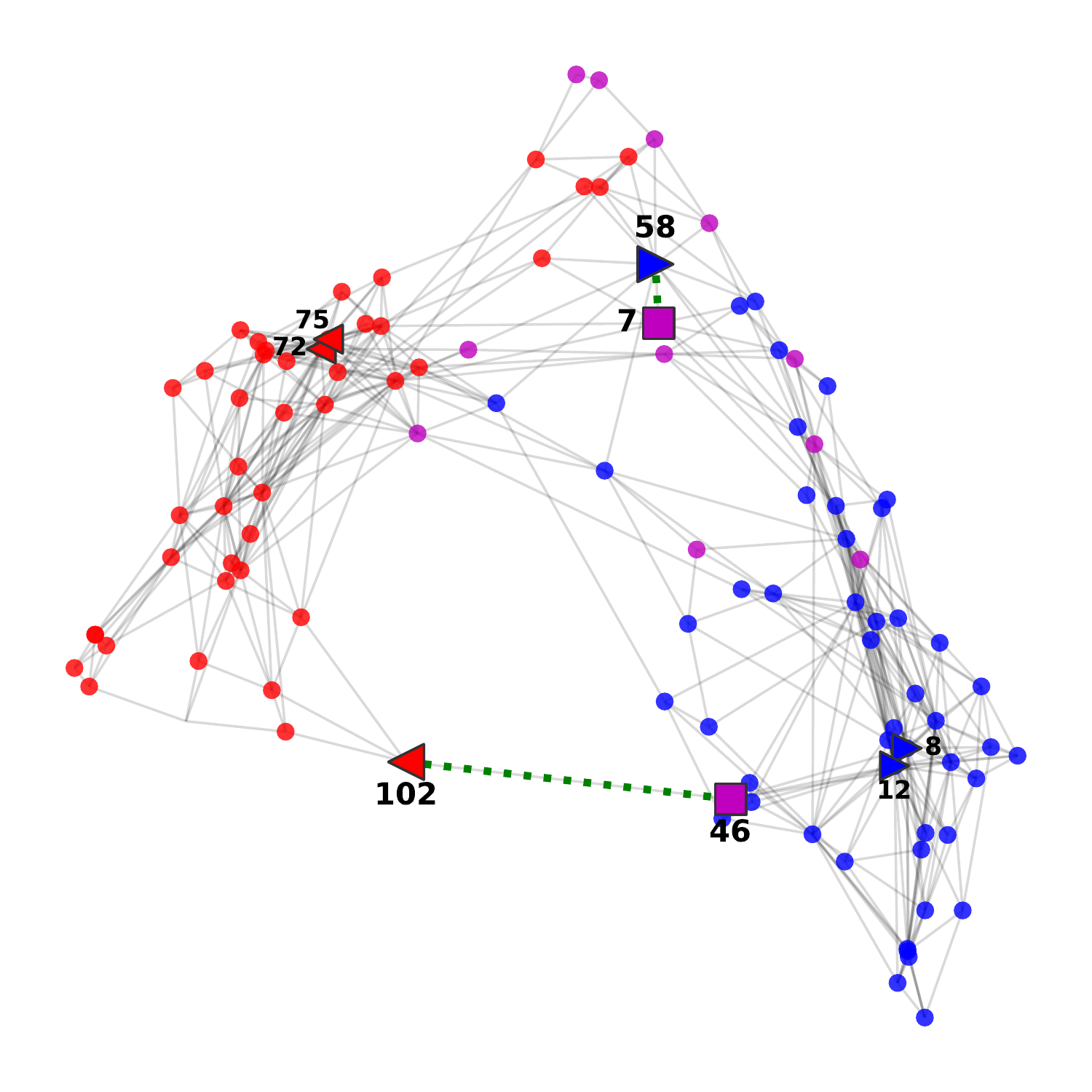}
\vspace{-5mm}
\caption{Case study on Polbooks with the most and least sensitive edge deletion. \label{fig:case_polbooks_e}}
\end{figure}

The edge deletions are marked in Fig.~\ref{fig:case_polbooks_e}, and we see the most sensitive ones are cross-community while the least sensitive ones are within-community.
Similar to the Karate case, node pair $(46, 102)$ has each other as the only cross-community link, after deleting which the node embeddings will be affected significantly.
Edge $(7, 58)$ is in the boundary between liberal and conservative nodes, and it has a neutral book.
As the predictions in the boundary are already uncertain, one edge deletion would fluctuate many predictions, resulting in high sensitivity.
The least-sensitive edge deletions are not only within-community, but are also between high-degree nodes, i.e.,  $d_{72} = 22$, $d_{75} = 16$, $d_{8} = d_{12} = 25$.
These nodes have already been well connected to nodes of the same type, thus they have stable embeddings and the deletions have little influence on relevant predictions.

\begin{figure}[hbt!]
\centering
\includegraphics[width=\columnwidth]{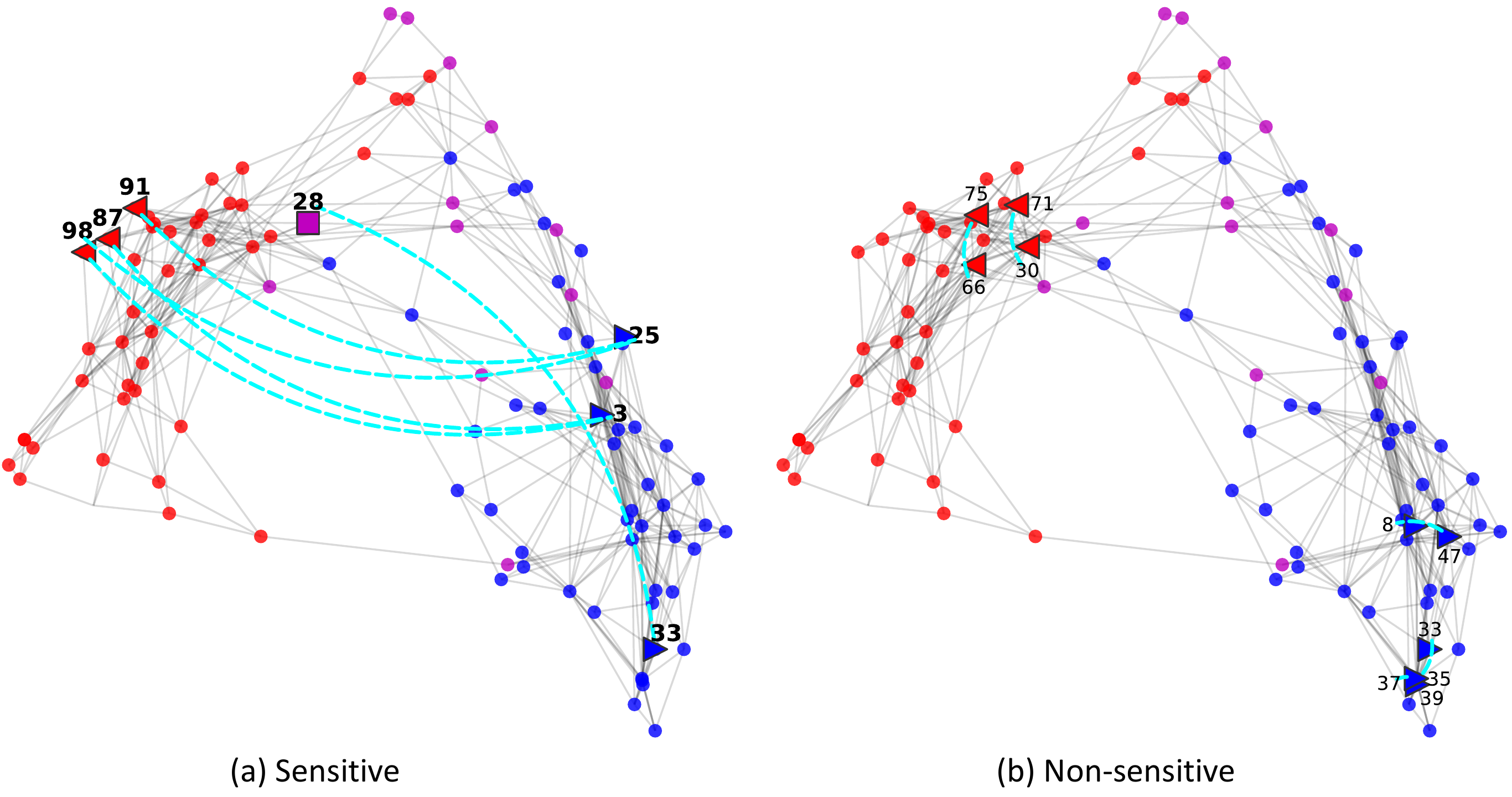}
\vspace{-7mm}
\caption{Case study on Polbooks with the most and least sensitive edge addition. \label{fig:case_polbooks_ne}}
\end{figure}

We mark the edge additions separately for the sensitive and non-sensitive perturbations in Fig.~\ref{fig:case_polbooks_ne}, to contrast their difference.
The left Fig.~\ref{fig:case_polbooks_ne} (a) shows the top 5 sensitive edge additions are all cross-community, and all include at least one node at the distant place from the opposing community, i.e., nodes $33$, $91$, $87$, $98$.
Being distant means those nodes have only the within-community connections, while adding a cross-community link would confuse the link predictor on the predictions for many relevant node pairs.
Meanwhile, as the sensitive perturbations involve low-degree nodes, they are usually unnoticeable while weighted highly by those nodes.
The non-sensitive edge additions are similar to the non-sensitive deletions in the sense that both have the pair of nodes embedded closely.
As long as the two nodes are mapped closely in the embedding space, it makes little difference if they are connected and the node degree does not matter much.

Interestingly, our observations in the case studies agree only partially with a heuristic community detection attack strategy called DICE~\cite{waniek2018hiding},
which has been used as a baseline for attacking link prediction in~\cite{chen2020link}.
Inspired by modularity, DICE randomly disconnect internally and connect externally~\cite{waniek2018hiding}, of which the goal is to hide the a group of nodes from being detected as a community.
Our analysis agrees with connecting externally, while for link prediction the disconnection should also be {\it external}, meaning that disconnecting internally might not work for link prediction.
If the internal disconnection are sampled to node pairs that are closely positioned, the attack will have the little influence.
Therefore, it might not be suitable to use DICE for link prediction attacks.

\subsection{Quality and Runtime of Approximations}
We use the sensitivity measured by re-embedding (RE) as the ground truth impact of the small perturbations. 
The quality of an approximation is determined by how close it is to the ground truth.
As the sensitivity is a ranked measure, we use the normalized discounted cumulative gain (NDCG) to evaluate the quality of the empirical approximation IPRE and the theoretical approximation with the diagonal Hessian blocks Approx.
The closer the NDCG value is to $1$, the better.
We do not include the theoretical approximation with the exact Hessian because it can be more computationally expensive than RE for large networks.
To show the significance, the p-value of each NDCG is found with randomization test of 1,000 samples.
The runtime for computing the sensitivity of one edge flip is recorded on a server with Intel Xeon Gold CPU 3.00GHz and 1024GB RAM.

Shown in Table~\ref{tab:ndcg} are the quality of the approximations on five real-world networks.
The first two columns show how well IPRE and Approx approximate RE, and the third column shows how well Approx approximates IPRE.
We see the NDCG values in the table are all significantly high.
Comparing to Approx, IPRE better approximates RE, and as the network size gets relatively large, the NDCG is alway larger than $0.99$, indicating that the larger the network, the more local the impact of a small perturbation.
For Approx, the NDCG for approximating RE are high across datasets, but it is even higher for IPRE.
The reason is that both Approx and IPRE essentially make the same assumption that the influence of the perturbation will be spread only to the one-hop neighborhood.

The approximations are not only effective, but also time-efficient.
We see in Table~\ref{tab:runtime} that RE is the slowest, IPRE is faster, and Approx is significantly much faster than the previous two empirical approaches, especially for larger networks.
On the Polblogs network, Approx is 36k times faster than RE and 22k times faster than IPRE.
It shows that our method also scales to large networks.

\begin{table}[hbt!]
\centering
\caption{Quality of the Approximations - NDCG}
\vspace{-2mm}
\label{tab:ndcg}
\begin{tabular}{|c|c|c|c|c|c|c|c|}
\hline
    ground truth      & \multicolumn{4}{c|}{RE}                                 &  & \multicolumn{2}{c|}{IPRE}   \\ \hline
   approximation       & \multicolumn{2}{c|}{IPRE} & \multicolumn{2}{c|}{Approx} &  & \multicolumn{2}{c|}{Approx} \\ \hline
          & NDCG        & p-value     & NDCG         & p-value      &  & NDCG         & p-value      \\ \hline
Polbooks ($n=105$) & 0.9691      & 0.0         & 0.9700       & 0.0          &  & 0.9873       & 0.0          \\ \hline
C.elegans ($n=297$) & 0.9977      & 0.0         & 0.9880       & 0.0          &  & 0.9905       & 0.0          \\ \hline
USAir ($n=332$)    & 0.9902      & 0.0         & 0.9697       & 0.0          &  & 0.9771       & 0.0          \\ \hline
MP   ($n=567$)    & 0.9985      & 0.0         & 0.9961       & 0.0          &  & 0.9960       & 0.0          \\ \hline
Polblogs ($n=1222$) & 0.9962      & 0.0         & 0.9897       & 0.0          &  & 0.9899       & 0.0          \\ \hline
\end{tabular}
\end{table}

\section{Conclusion \label{sec:conclusion}}
In this work we study the adversarial robustness of a probabilistic network embedding model CNE for the link prediction task by measuring the sensitivity of the link predictions of the model to small adversarial perturbations of the network.
Our approach allows us to identify the most vulnerable links and non-links that if perturbed will have large impact on the model's link prediction performance, which can be used for further investigation, such as defending attacks by protecting those.
With two case studies, we analyze the characteristics of the most and least sensitive perturbations for link prediction with CNE.
Then we empirically confirm that our theoretical approximation of the sensitivity measure is both effective and efficient, 
meaning that the worst-case perturbations for link prediction using CNE can be identified successfully in a time-efficient manner with our method.
For future work, we plan to explore the potential of our theoretical approximation to construct a regularizer for adversarially robust network embedding or to develop robustness certificates for link prediction.

\section*{Acknowledgement}
The research leading to these results has received funding from the European Research Council under the European Union's Seventh Framework Programme (FP7/2007-2013) (ERC Grant Agreement no. 615517), and under the European Union's Horizon 2020 research and innovation programme (ERC Grant Agreement no. 963924), from the Flemish Government under the ``Onderzoeksprogramma Artifici{\"e}le Intelligentie (AI) Vlaanderen'' programme, and from the FWO (project no. G091017N, G0F9816N, 3G042220).

 \bibliographystyle{splncs04}
 \bibliography{mybib}

\begin{thebibliography}{10}
\providecommand{\url}[1]{\texttt{#1}}
\providecommand{\urlprefix}{URL }
\providecommand{\doi}[1]{https://doi.org/#1}

\bibitem{adamic2005political}
Adamic, L.A., Glance, N.: {The Political Blogosphere and the 2004 U.S.
  Election: Divided They Blog}. In: Proc. of LinkKDD 2005. pp. 36--43 (2005)

\bibitem{bojchevski2019adversarial}
Bojchevski, A., G{\"u}nnemann, S.: {Adversarial Attacks on Node Embeddings via
  Graph Poisoning}. In: Proc. of the 36th ICML. pp. 695--704 (2019)

\bibitem{bojchevski2019certifiable}
Bojchevski, A., G\"{u}nnemann, S.: {Certifiable Robustness to Graph
  Perturbations}. In: Proc. of the 33rd NeurIPS. vol.~32 (2019)

\bibitem{chen2020link}
Chen, J., Lin, X., Shi, Z., Liu, Y.: {Link Prediction Adversarial Attack via
  Iterative Gradient Attack}. IEEE Trans. Comput. Soc. Syst.  \textbf{7}(4),
  1081--1094 (2020)

\bibitem{chen2020survey}
Chen, L., Li, J., Peng, J., Xie, T., Cao, Z., Xu, K., He, X., Zheng, Z.: {A
  Survey of Adversarial Learning on Graphs}. arXiv preprint arXiv:2003.05730
  (2020)

\bibitem{chen2021alpine}
Chen, X., Kang, B., Lijffijt, J., De~Bie, T.: {ALPINE: Active Link Prediction
  Using Network Embedding}. Applied Sciences  \textbf{11}(11), ~5043 (2021)

\bibitem{dai2018adversarial}
Dai, H., Li, H., Tian, T., Huang, X., Wang, L., Zhu, J., Song, L.: Adversarial
  attack on graph structured data. In: Proc. of the 35th ICML. pp. 1115--1124
  (2018)

\bibitem{dai2018adversarialNE}
Dai, Q., Li, Q., Tang, J., Wang, D.: {Adversarial Network Embedding}. In: Proc.
  of the 32nd AAAI. vol.~32 (2018)

\bibitem{dai2019adversarial}
Dai, Q., Shen, X., Zhang, L., Li, Q., Wang, D.: {Adversarial Training Methods
  for Network Embedding}. In: Proc. of the 28th WWW. pp. 329--339 (2019)

\bibitem{fard2015neighborhood}
Fard, A.M., Wang, K.: {Neighborhood Randomization for Link Privacy in Social
  Network Analysis}. World Wide Web  \textbf{18}(1),  9--32 (2015)

\bibitem{feng2019graph}
{Feng, Fuli and He, Xiangnan and Tang, Jie and Chua, Tat-Seng}: Graph
  adversarial training: Dynamically regularizing based on graph structure. IEEE
  Trans. Knowl. Data Eng.  \textbf{33}(6),  2493--2504 (2021)

\bibitem{gao2020certified}
{Gao, Zhidong and Hu, Rui and Gong, Yanmin}: {Certified Robustness of Graph
  Classification against Topology Attack with Randomized Smoothing}. In: Proc.
  of the GLOBECOM 2020. pp.~1--6 (2020)

\bibitem{goodfellow2014explaining}
Goodfellow, I.J., Shlens, J., Szegedy, C.: {Explaining and Harnessing
  Adversarial Examples}. In: Proc. of the 3rd ICLR (2015)

\bibitem{gori2005new}
Gori, M., Monfardini, G., Scarselli, F.: {A New Model for Learning in Graph
  Domains}. In: Proc. of 2005 IEEE IJCNN. vol.~2, pp. 729--734 (2005)

\bibitem{grover2016node2vec}
Grover, A., Leskovec, J.: {node2vec: Scalable Feature Learning for Networks}.
  In: Proc. of the 22nd ACM SIGKDD. pp. 855--864 (2016)

\bibitem{hamilton2017inductive}
Hamilton, W., Ying, Z., Leskovec, J.: {Inductive Representation Learning on
  Large Graphs}. In: Proc. of the 31st NeurIPS. vol.~30 (2017)

\bibitem{handcock2003statnet}
Handcock, M.S., Hunter, D.R., Butts, C.T., Goodreau, S.M., Morris, M.:
  {statnet: An R package for the Statistical Modeling of Social Networks}. Web
  page http://www.csde.washington.edu/statnet  (2003)

\bibitem{ieee1990ieee}
IEEE: {IEEE Standard Glossary of Software Engineering Terminology}. IEEE Std
  610.12-1990 pp. 1--84 (1990). \doi{10.1109/IEEESTD.1990.101064}

\bibitem{jia2020certified}
Jia, J., Wang, B., Cao, X., Gong, N.Z.: {Certified Robustness of Community
  Detection against Adversarial Structural Perturbation via Randomized
  Smoothing}. In: Proc. of the 29th WWW. pp. 2718--2724 (2020)

\bibitem{jin2020certified}
Jin, H., Shi, Z., Peruri, V.J.S.A., Zhang, X.: {Certified Robustness of Graph
  Convolution Networks for Graph Classification under Topological Attacks}. In:
  Proc. of the 34th NeurIPS. vol.~33, pp. 8463--8474 (2020)

\bibitem{jin2020adversarial}
Jin, W., Li, Y., Xu, H., Wang, Y., Tang, J.: {Adversarial attacks and defenses
  on graphs: A review and empirical study}. arXiv preprint arXiv:2003.00653
  (2020)

\bibitem{kang2018conditional}
Kang, B., Lijffijt, J., De~Bie, T.: {Conditional Network Embeddings}. In: Proc.
  of the 7th ICLR (2019)

\bibitem{kang2019explaine}
Kang, B., Lijffijt, J., De~Bie, T.: {ExplaiNE: An Approach for Explaining
  Network Embedding-based Link Predictions}. arXiv preprint arXiv:1904.12694
  (2019)

\bibitem{kipf2017semi}
Kipf, T.N., Welling, M.: {Semi-Supervised Classification with Graph
  Convolutional Networks}. In: Proc. of the 5th ICLR (2017)

\bibitem{liben2007link}
Liben-Nowell, D., Kleinberg, J.: {The Link-Prediction Problem for Social
  Networks}. J. Am. Soc. Inf. Sci. Technol.  \textbf{58}(7),  1019--1031 (2007)

\bibitem{lin2020adversarial}
Lin, W., Ji, S., Li, B.: {Adversarial Attacks on Link Prediction Algorithms
  Based on Graph Neural Networks}. In: Proc. of the 15th ACM AsiaCCS. pp.
  370--380 (2020)

\bibitem{liunetwork}
Liu, X., Tang, J.: {Network Representation Learning: A Macro and Micro View}

\bibitem{liu2021adversarial}
Liu, Z., Larson, M.: {Adversarial Item Promotion: Vulnerabilities at the Core
  of Top-N Recommenders That Use Images to Address Cold Start}. In: Proc. of
  the 30th WWW. pp. 3590--3602 (2021)

\bibitem{ma2019attacking}
Ma, Y., Wang, S., Derr, T., Wu, L., Tang, J.: {Attacking Graph Convolutional
  Networks via Rewiring}. arXiv preprint arXiv:1906.03750  (2019)

\bibitem{mara2020benchmarking}
Mara, A.C., Lijffijt, J., De~Bie, T.: {Benchmarking Network Embedding Models
  for Link Prediction: Are We Making Progress?} In: Proc. of the 7th IEEE DSAA.
  pp. 138--147 (2020)

\bibitem{martinez2016survey}
Mart{\'\i}nez, V., Berzal, F., Cubero, J.C.: {A Survey of Link Prediction in
  Complex Networks}. ACM Comput. Surv.  \textbf{49}(4),  1--33 (2016)

\bibitem{mirzasoleiman2020coresets}
Mirzasoleiman, B., Cao, K., Leskovec, J.: {Coresets for Robust Training of Deep
  Neural Networks against Noisy Labels}. In: Proc. of the 34th NeurIPS.
  vol.~33, pp. 11465--11477 (2020)

\bibitem{miyato2018virtual}
Miyato, T., Maeda, S.i., Koyama, M., Ishii, S.: {Virtual Adversarial Training:
  A Regularization Method for Supervised and Semi-Supervised Learning}. IEEE
  PAMI  \textbf{41}(8),  1979--1993 (2018)

\bibitem{pan2018adversarially}
Pan, S., Hu, R., Long, G., Jiang, J., Yao, L., Zhang, C.: {Adversarially
  Regularized Graph Autoencoder for Graph Embedding}. In: Proc. of the 27th
  IJCAI. pp. 2609--2615 (2018)

\bibitem{perozzi2014deepwalk}
Perozzi, B., Al-Rfou, R., Skiena, S.: {DeepWalk: Online Learning of Social
  Representations}. In: Proc. of the 20th ACM SIGKDD. pp. 701--710 (2014)

\bibitem{qiu2018network}
Qiu, J., Dong, Y., Ma, H., Li, J., Wang, K., Tang, J.: {Network Embedding as
  Matrix Factorization: Unifying DeepWalk, LINE, PTE, and node2vec}. In: Proc.
  of the 11th ACM WSDM. pp. 459--467 (2018)

\bibitem{sun2018adversarial}
Sun, L., Dou, Y., Yang, C., Wang, J., Yu, P.S., He, L., Li, B.: {Adversarial
  attack and defense on graph data: A survey}. arXiv preprint arXiv:1812.10528
  (2018)

\bibitem{sun2018data}
Sun, M., Tang, J., Li, H., Li, B., Xiao, C., Chen, Y., Song, D.: {Data
  poisoning attack against unsupervised node embedding methods}. arXiv preprint
  arXiv:1810.12881  (2018)

\bibitem{tang2015line}
Tang, J., Qu, M., Wang, M., Zhang, M., Yan, J., Mei, Q.: {LINE}: Large-scale
  information network embedding. In: Proc. of the 24th WWW. pp. 1067--1077
  (2015)

\bibitem{tang2020transferring}
Tang, X., Li, Y., Sun, Y., Yao, H., Mitra, P., Wang, S.: {Transferring
  Robustness for Graph Neural Network against Poisoning Attacks}. In: Proc. of
  the 13th WSDM. pp. 600--608 (2020)

\bibitem{wang2017community}
Wang, X., Cui, P., Wang, J., Pei, J., Zhu, W., Yang, S.: {Community Preserving
  Network Embedding}. In: Proc. of the 31st AAAI. vol.~31 (2017)

\bibitem{waniek2018hiding}
Waniek, M., Michalak, T.P., Wooldridge, M.J., Rahwan, T.: Hiding individuals
  and communities in a social network. Nature Human Behaviour  \textbf{2}(2),
  139--147 (2018)

\bibitem{waniek2019hide}
Waniek, M., Zhou, K., Vorobeychik, Y., Moro, E., Michalak, T.P., Rahwan, T.:
  {How to Hide One's Relationships from Link Prediction Algorithms}. Scientific
  Reports  \textbf{9}(1),  1--10 (2019)

\bibitem{watts1998collective}
Watts, D.J., Strogatz, S.H.: Collective dynamics of `small-world' networks.
  Nature  \textbf{393}(6684),  440--442 (1998)

\bibitem{wu2019adversarial}
Wu, H., Wang, C., Tyshetskiy, Y., Docherty, A., Lu, K., Zhu, L.: {Adversarial
  Examples for Graph Data: Deep Insights into Attack and Defense}. In: Proc. of
  the 28th IJCAI. pp. 4816--4823 (2019)

\bibitem{xu2019topology}
Xu, K., Chen, H., Liu, S., Chen, P.Y., Weng, T.W., Hong, M., Lin, X.: {Topology
  Attack and Defense for Graph Neural Networks: An Optimization Perspective}.
  In: Proc. of the 28th IJCAI. pp. 3961--3967 (2019)

\bibitem{yang2017fake}
Yang, G., Gong, N.Z., Cai, Y.: {Fake Co-visitation Injection Attacks to
  Recommender Systems}. In: Proc. of the 24th NDSS (2017)

\bibitem{yu2019target}
Yu, S., Zhao, M., Fu, C., Zheng, J., Huang, H., Shu, X., Xuan, Q., Chen, G.:
  {Target Defense Against Link-Prediction-Based Attacks via Evolutionary
  Perturbations}. IEEE Trans. Knowl. Data Eng.  \textbf{33}(2),  754--767
  (2021)

\bibitem{zachary1977information}
Zachary, W.W.: {An Information Flow Model for Conflict and Fission in Small
  Groups}. J. Anthropol. Res.  \textbf{33}(4),  452--473 (1977)

\bibitem{zhang2020practical}
Zhang, H., Li, Y., Ding, B., Gao, J.: {Practical Data Poisoning Attack against
  Next-Item Recommendation}. In: Proc. of the 29th WWW. pp. 2458--2464 (2020)

\bibitem{zhang2018link}
Zhang, M., Chen, Y.: {Link Prediction Based on Graph Neural Networks}. In:
  Proc. of the 32nd NeurIPS. vol.~31 (2018)

\bibitem{zheng2020robust}
Zheng, C., Zong, B., Cheng, W., Song, D., Ni, J., Yu, W., Chen, H., Wang, W.:
  {Robust Graph Representation Learning via Neural Sparsification}. In: Proc.
  of the 37th ICML. pp. 11458--11468 (2020)

\bibitem{zhou2019adversarial}
Zhou, K., Michalak, T.P., Vorobeychik, Y.: {Adversarial Robustness of
  Similarity-based Link Prediction}. In: Proc. of the 19th IEEE ICDM. pp.
  926--935 (2019)

\bibitem{zhou2019attacking}
Zhou, K., Michalak, T.P., Waniek, M., Rahwan, T., Vorobeychik, Y.: {Attacking
  Similarity-Based Link Prediction in Social Networks}. In: Proc. of the 18th
  AAMAS. pp. 305--313 (2019)

\bibitem{zhu2019robust}
Zhu, D., Zhang, Z., Cui, P., Zhu, W.: {Robust graph convolutional networks
  against adversarial attacks}. In: Proc. of the 25th ACM SIGKDD. pp.
  1399--1407 (2019)

\bibitem{zugner2018adversarial}
Z{\"u}gner, D., Akbarnejad, A., G{\"u}nnemann, S.: {Adversarial attacks on
  neural networks for graph data}. In: Proc. of the 24th ACM SIGKDD. pp.
  2847--2856 (2018)

\bibitem{zugner2020adversarial}
Z{\"u}gner, D., Borchert, O., Akbarnejad, A., Guennemann, S.: {Adversarial
  Attacks on Graph Neural Networks: Perturbations and their Patterns}. ACM
  Trans. Knowl. Discov. Data  \textbf{14}(5),  1--31 (2020)

\bibitem{zugner2019adversarial}
Z{\"u}gner, D., G{\"u}nnemann, S.: {Adversarial Attacks on Graph Neural
  Networks via Meta Learning}. In: Proc. of the 7th ICLR (2019)

\bibitem{zugner2019certifiable}
Z{\"u}gner, D., G{\"u}nnemann, S.: {Certifiable robustness and robust training
  for graph convolutional networks}. In: Proc. of the 25th ACM SIGKDD. pp.
  246--256 (2019)

\bibitem{zugner2020certifiable}
Z{\"u}gner, D., G{\"u}nnemann, S.: {Certifiable robustness of graph
  convolutional networks under structure perturbations}. In: Proc. of the 26th
  ACM SIGKDD. pp. 1656--1665 (2020)

\end{thebibliography}
 
 \end{document}